\documentclass[12pt]{elsarticle}
\usepackage{amsmath}
\usepackage{amssymb}
\usepackage{url}
\usepackage{graphics}

\newcommand{\ang}[1]{\langle#1\rangle}

\newcommand{\T}{{\rm T}}

\newcommand{\xvec}[1]{\ifcase 3{#1} {\ang {x_1,x_2,x_3} } \else
\ifcase 4{#1} {\ang{x_1,x_2,x_3,x_4}} \else {\ang {x_1,\ldots,x_{#1}}}\fi\fi}
\newcommand{\yvec}[1]{\ifcase 3{#1} {\ang {y_1,y_2,y_3} } \else
\ifcase 4{#1} {\ang{y_1,y_2,y_3,y_4}} \else {\ang {y_1,\ldots,y_{#1}}}\fi\fi}
\newcommand{\zvec}[1]{\ifcase 3{#1} {\ang {z_1,z_2,z_3} } \else
\ifcase 4{#1} {\ang{z_1,z_2,z_3,z_4}} \else {\ang {z_1,\ldots,z_{#1}}}\fi\fi}
\newcommand{\vecc}[2]{\ifcase 3{#2} {\ang { {#1}_1,{#1}_2,{#1}_3 } } \else
\ifcase 4{#1} {\ang { {#1}_1,{#1}_2,{#1}_3,{#1}_{4} } }
\else {\ang { {#1}_1,\ldots,{#1}_{#2}}}\fi\fi}
\newcommand{\veccd}[3]{\ifcase 3{#2} {\ang { {#1}_{{#3}1},{#1}_{{#3}2},{#1}_{{#3}3} } } \else
\ifcase 4{#1} {\ang { {#1}_{{#3}1},{#1}_{{#3}2},{#1}_{#3}3},{#1}_{{#3}4} }
\else {\ang { {#1}_{{#3}1},\ldots,{#1}_{{#3}{#2}}}}\fi\fi}
\newcommand{\veccz}[2]{\ifcase 3{#2} {\ang { {#1}_0,{#1}_2,{#1}_3 } } \else
\ifcase 4{#1} {\ang { {#1}_0,{#1}_2,{#1}_3,{#1}_{4} } }
\else {\ang { {#1}_0,\ldots,{#1}_{#2}}}\fi\fi}
\newcommand{\xve}[1]{\ifcase 3{#1} {x_1,x_2,x_3} \else
\ifcase 4{#1} {x_1,x_2,x_3,x_4} \else {x_1,\ldots,x_{#1}}\fi\fi}
\newcommand{\yve}[1]{\ifcase 3{#1} {y_1,y_2,y_3} \else
\ifcase 4{#1} {y_1,y_2,y_3,y_4} \else {y_1,\ldots,y_{#1}}\fi\fi}
\newcommand{\zve}[1]{\ifcase 3{#1} {z_1,z_2,z_3} \else
\ifcase 4{#1} {z_1,z_2,z_3,z_4} \else {z_1,\ldots,z_{#1}}\fi\fi}
\newcommand{\ve}[2]{\ifcase 3#2 {{#1}_1,{#1}_2,{#1}_3} \else
\ifcase 4#2 {{#1}_1,{#1}_2,{#1}_3,{#1}_{4}}
\else {{#1}_1,\ldots,{#1}_{#2}}\fi\fi}
\newcommand{\ved}[3]{\ifcase 3#2 {{#1}_{{#3}1},{#1}_{{#3}2},{#1}_{{#3}3}} \else
\ifcase 4#2 {{#1}_{{#3}1},{#1}_{{#3}2},{#1}_{{#3}3},{#1}_{{#3}4}}
\else {{#1}_{{#3}1},\ldots,{#1}_{{#3}{#2}}}\fi\fi}
\newcommand{\fuve}[3]{
\ifcase 3#2
{{#3}({#1}_1),{#3}({#1}_2,{#3}({#1}_3)} \else
\ifcase 4#2
{{#3}({#1}_1),{#3}({#1}_2),{#3}({#1}_3),{#3}({#1}_4)}
\else
{{#3}({#1}_1),\ldots,{#3}({#1}_{#2})}\fi\fi}
\newcommand{\setmathchar}[1]{\ifmmode#1\else$#1$\fi}
\newcommand{\vlist}[2]{%
	\setmathchar{%
		\compound#2\one{#2}\two
		\ifcompound
			({#1}_1,\ldots,{#1}_{#2})
		\else
			\ifcat N#2
				({#1}_1,\ldots,{#1}_{#2})
			\else
				\ifcase#2
					({#1}_0)\or
					({#1}_1)\or
					({#1}_1,{#1}_2)\or
					({#1}_1,{#1}_2,{#1}_3)\or
					({#1}_1,{#1}_2,{#1}_3,{#1}_4)\else
					({#1}_1,\ldots,{#1}_{#2})
				\fi
			\fi
		\fi}}

\newif\ifcompound
\def\compound#1\one#2\two{%
	\def\one{#1}
	\def\two{#2}
	\if\one\two
		\compoundfalse
	\else
		\compoundtrue
	\fi}

\newcommand{\xwe}[1]{\ifcase 3{#1} {x_1\wedge x_2\wedge x_3} \else
\ifcase 4{#1} {x_1\wedge x_2\wedge x_3\wedge x_4} \else {x_1\wedge \cdots \wedge
x_{#1}}\fi\fi}
\newcommand{\we}[2]{\ifcase 3#2 {\ang { {#1}_1\wedge {#1}_2\wedge {#1}_3 } } \else
\ifcase 4{#1} {\ang { {#1}_1\wedge {#1}_2\wedge {#1}_3\wedge {#1}_{4} } }
\else {\ang { {#1}_1\wedge \cdots\wedge {#1}_{#2}}}\fi\fi}
\newcommand{\s}[1]{\s_{#1}}

\newcommand{\monus}{\;\raise.5ex\hbox{{${\buildrel
    \ldotp\over{\hbox to 6pt{\hrulefill}}}$}}\;}

\newcounter{savenumi}

\newtheorem{theoremfoo}{Theorem}[section] 
\newenvironment{theorem}{\pagebreak[1]\begin{theoremfoo}}{\end{theoremfoo}}

\newtheorem{lemmafoo}[theoremfoo]{Lemma}

\newtheorem{conjecturefoo}[theoremfoo]{Conjecture}

\newenvironment{conjecture}{\pagebreak[1]\begin{conjecturefoo}}{\end{conjecturefoo}}

\newtheorem{conventionfoo}[theoremfoo]{Convention}

\newtheorem{porismfoo}[theoremfoo]{Porism}

\newtheorem{gamefoo}[theoremfoo]{Game}

\newtheorem{corollaryfoo}[theoremfoo]{Corollary}

\newtheorem{claimfoo}[theoremfoo]{Claim}

\newtheorem{openfoo}[theoremfoo]{Open Problem}

\newtheorem{exercisefoo}{Exercise}

\newcommand{\fig}[1] 
{
 \begin{figure}
 \begin{center}
 \input{#1}
 \end{center}
 \end{figure}
}

\newtheorem{potanafoo}[theoremfoo]{Potential Analogue}

\newtheorem{notefoo}[theoremfoo]{Note}

\newtheorem{notabenefoo}[theoremfoo]{Nota Bene}

\newtheorem{nttn}[theoremfoo]{Notation}

\newtheorem{empttn}[theoremfoo]{Empirical Note}

\newtheorem{examfoo}[theoremfoo]{Example}

\newtheorem{dfntn}[theoremfoo]{Def}
\newenvironment{definition}{\pagebreak[1]\begin{dfntn}\rm}{\end{dfntn}}

\newtheorem{propositionfoo}[theoremfoo]{Proposition}

\newenvironment{proof}
    {\pagebreak[1]{\narrower\noindent {\bf Proof:\quad\nopagebreak}}}{\QED}

\newcommand{\yyskip}{\penalty-50\vskip 5pt plus 3pt minus 2pt}
\newcommand{\blackslug}{\hbox{\hskip 1pt
        \vrule width 4pt height 8pt depth 1.5pt\hskip 1pt}}
\newcommand{\QED}{{\penalty10000\parindent 0pt\penalty10000
        \hskip 8 pt\nolinebreak\blackslug\hfill\lower 8.5pt\null}
        \par\yyskip\pagebreak[1]}

\newcommand{\BBB}{{\penalty10000\parindent 0pt\penalty10000
        \hskip 8 pt\nolinebreak\hbox{\ }\hfill\lower 8.5pt\null}
        \par\yyskip\pagebreak[1]}

\newtheorem{factfoo}[theoremfoo]{Fact}
\newenvironment{fact}{\pagebreak[1]\begin{factfoo}}{\end{factfoo}}

\newenvironment{block}{\begin{list}{\hbox{}}{\leftmargin 1em
    \itemindent -1em \topsep 0pt \itemsep 0pt \partopsep 0pt}}{\end{list}}

\journal{Theoretical Computer Science}

\begin{document}
\begin{frontmatter}

\title{Speedup for Natural Problems and Noncomputability\tnoteref{t1}}
\tnotetext[t1]{This paper is in honor of the retirement of Benjamin Klein from Davidson College. The views expressed in this column are those of the author and should not be attributed to the International Monetary Fund, its Executive Board, or its management. This paper could not have been prepared without encouragement from Marius Zimand and Bill Gasarch. This paper arose from a suggestion by Richard Beigel. I would also like to thank an anonymous referee, Scott Aaronson, Amir Ben-Amram, Neil Christensen, Lance Fortnow, J\"{o}rg Flum, Yijia Chen, and participants in a University of Maryland Complexity Seminar who provided useful comments. Remaining errors are my own.}
\author{Hunter Monroe}
\ead{hmonroe.tcs@huntermonroe.com}
\address{International Monetary Fund, 700 19th St, NW, Washington, DC 20431}
\date{\today}
\begin{abstract}
A resource-bounded version of the statement ``no algorithm recognizes all non-halting Turing machines'' is equivalent to an  infinitely often (i.o.) superpolynomial speedup for the time required to accept any (paddable) $\textbf{coNP}$-complete language and also equivalent to a superpolynomial speedup in proof length in propositional proof systems for tautologies, each of which implies $\textbf{P}\neq \textbf{NP}$. This suggests a correspondence between the properties ``has no algorithm at all'' and ``has no best algorithm'' which seems relevant to open problems in computational and proof complexity.\end{abstract}



\end{frontmatter}
\section{Introduction}
Informally, a language $L$ has speedup if, for any Turing machine (TM) for $L$, there exists one that is better. Blum~\cite{Blum1967} exhibited languages that have almost-everywhere speedup, which are unnatural being constructed solely for that purpose. The possibility of weaker speedups for natural languages has received less attention \cite{Monroe:2008}. Some suspect that integer multiplication and matrix multiplication (MM) have a slight, superlinear speedup \cite{SchnorrStumpf,Meyer,Blum2007}, reflecting in part the large number of algorithms for these problems---about 13 and 18 respectively \cite{Bernstein,Pan}. In fact, there is no best Strassen-style bilinear MM identity \cite{Coppersmith82}.

We identify an intuitive condition which, like several others in the literature, is equivalent to an  infinitely often (i.o.) superpolynomial speedup for the time required to accept any (paddable) $\textbf{coNP}$-complete language and also equivalent to a superpolynomial speedup in proof length in propositional proof systems for tautologies, each of which implies $\textbf{P}\neq \textbf{NP}$. This condition is a resource-bounded version of the statement ``no algorithm recognizes all non-halting TMs'', suggesting a correspondence between the properties ``has no algorithm at all'' and ``has no best algorithm'' which seems relevant to open problems in computational and proof complexity.
\section{Speedup for \textbf{coNP}-Complete Languages}\label{superpolysection}
Consider this well-known fact from computability theory:
\begin{fact}\label{factnonhalting}
Given any TM $M$ accepting only ``non-halting'' $\langle N,x\rangle$ for which TM $N$ does not halt on input $x$, $M$ fails to accept some particular non-halting $\langle N',x'\rangle$. In other words, the set of non-halting $\langle N,x\rangle$ is not computably enumerable (c.e.).
\end{fact}
By implication, there is a better TM $M'$ that correctly accepts more non-halting inputs than $M$ by accepting the input $\langle N',x'\rangle$ and otherwise running $M$. This section considers a corresponding resource-bounded statement in complexity theory regarding $N$ which do not halt on $x$ within $t$ steps.

\textbf{Notation:} $M$ and $M'$ will denote deterministic TMs throughout the paper, and, henceforth, $N$ and $N'$ will denote nondeterministic TMs. Define $\texttt{BHP}=\{\langle N,x,1^t\rangle|$ there is at least one accepting path of nondeterministic TM $N$
on input $x$ with $t$ or fewer steps$\}$ and define $\texttt{coBHP}=\{\langle N,x,1^t\rangle|\langle N,x,1^t\rangle\notin \texttt{BHP}\}$.
If $M$ is a deterministic TM, then $\T_M$ is the function that maps a string $x$ to how many steps $M(x)$ takes. Say that $M$ accepts a language $L$ if $M$ halts in an accepting state if and only if $x\in L$; $M$ may not halt on $x\notin L$. Say that $\langle N',x'\rangle$ is non-halting if $N'$ has no accepting path on input $x'$, in which case Fact \ref{factnonhalting} continues to hold for nondeterministic $N$. Note that $\texttt{BHP}$ is $\textbf{NP}$-complete with the accepting path of $N$ on $x$ as a certificate, and that $\texttt{coBHP}$ is $\textbf{coNP}$-complete.

The following condition corresponds to Fact \ref{factnonhalting}:
\begin{quote}
(*) For any $M$ accepting $\texttt{coBHP}$, there exists some non-halting $\langle N',x'\rangle$ such that the function $f(t)=T_M(N',x',1^t)$ is not bounded by any polynomial.\footnote{The function $f$ may depend on $M$, $N'$, and $x'$. For inputs not in $\texttt{coBHP}$, $M$ does not accept, but otherwise its behavior is not constrained.}
\end{quote}
Supposing $\textbf{P}\neq \textbf{NP}$ and therefore $\texttt{coBHP}\notin \textbf{P}$, condition (*) rules out the absurd possibility that some $M$ nevertheless can accept the subset of inputs beginning with any particular machine-input pair within a polynomial bound (for that subset). An intuition for why this condition might hold could be a belief that there is at least one $\langle N',x'\rangle$ for which $M$ must infinitely often use brute force to rule out all possible accepting paths of $N'$ on $x'$ with at most $t$ steps.\footnote{Condition (*) is equivalent to the statement that there is no $M$ deciding $\texttt{BHP}$ within time $O(t^{f(|N,x|)})$. Chen and Flum \cite{ChenFlum} show that under certain complexity theoretic assumptions, there is no such $M$ for $f$ computable.} Under (*), $\texttt{coBHP}$ has an i.o. superpolynomial speedup, defined as follows:\footnote{By contrast, Hirsch and Itsykson \cite{Hirsch} exhibit a $p$-optimal heuristic randomized algorithm for accepting the set of tautologies ($\texttt{TAUT}$), where the algorithm is allowed to accept non-tautologies erroneously with bounded probability. Levin \cite{Levin1973} exhibits a $p$-optimal witness search algorithm for any language in $\textbf{NP}$. Levin's algorithm dovetails every possible TM, runs any output produced through a predetermined witness verifier, and then prints out the first witness that is verified. However, even though $\texttt{SAT}\in\textbf{NP}$, K\"{o}bler and Messner \cite{Kobler} argue that accepting $\texttt{SAT}$ is likely to have superpolynomial speedup.}
\begin{definition}\label{defn}
For $M$ and $M'$ accepting a language $L$, write $M'\leq_p M$ if there exists a polynomial $p$ such that for all inputs $x\in L$,
\begin{equation}\label{inequality}
T_{M'}(x)\leq p(|x|,T_M(x)).
\end{equation}
If $M'\le_pM$ but it is not the case that $M\le_pM'$, write $M'<_pM$. If $L$ has a least element $M$ under $<_p$, say that $M$ is \emph{$p$-optimal} \cite{Krajicek} and otherwise that $L$ has \emph{(i.o.) superpolynomial speedup}.
\end{definition}
It is shown below that (*) is equivalent to a superpolynomial speedup for $\texttt{coBHP}$. This conclusion is significant, as superpolynomial speedup for accepting a particular (paddable) $\textbf{coNP}$-complete language is in fact equivalent to superpolynomial speedup for accepting any (paddable) $\textbf{coNP}$-complete language.\footnote{All known \textbf{coNP}-complete languages are paddable.} Furthermore, it is also equivalent to a superpolynomial speedup for proof length in propositional proof systems for the set of tautologies ($\texttt{TAUT}$) \cite{Krajicek}, defined as follows. A propositional proof system is a function $h\in \textbf{FP}$ with range $\texttt{TAUT}$ \cite{CookReckhow}. The proof system $h$ is $p$-optimal if for any other proof system $f$, there exists $g\in \textbf{FP}$ such that $h(g(x))=f(x)$ \cite{Krajicek}. Thus, (*) holds iff there is no $p$-optimal propositional proof system, so propositional proof systems have a superpolynomial speedup for proof length.
\begin{theorem}\label{superpolynomial}
The condition (*) holds if and only if $\verb"coBHP"$ has superpolynomial speedup.
\end{theorem}
\begin{proof}
$\Rightarrow$ Suppose condition (*) holds. Given $M$ accepting $\texttt{coBHP}$, choose $N',x'$ for $M$ in (*), so $f(t)=T_M(\langle N',x',1^t \rangle)$ is not polynomially bounded.
We create $M'$ as follows:
\begin{enumerate}
\item
Input $\langle N,x,1^t \rangle$.
\item
If $\langle N,x\rangle\ne \langle N',x'\rangle$ then run $M(\langle N,x,1^t\rangle)$.
\item
If $\langle N,x\rangle=\langle N',x'\rangle$ then accept immediately.
\end{enumerate}
Then $M'<_p M$, so $\texttt{coBHP}$ has superpolynomial speedup.

$\Leftarrow$ The converse follows from results of Chen and Flum;\footnote{If (*) does not hold, then $\texttt{coBHP}\in \textbf{XP}_{uni}$, where a parameterized problem $(Q,\kappa)$ is in $\textbf{XP}_{uni}$ if there is an $M$ deciding $x\in Q$ in time $|x|^{f(\kappa(x))}$ (for \texttt{coBHP}, $\langle N,x\rangle$ is the parameter $\kappa$). In that case, there is a $p$-optimal $M$ accepting any \textbf{coNP}-complete language, including \texttt{coBHP} (\cite{ChenFlum2} Theorem 8 and Lemma 18).} an anonymous referee proposed the following more direct argument. If (*) does not hold, it will be shown that there is a TM $M_{opt}$ which is $p$-optimal for \texttt{coBHP}. The strategy employed by this machine is to enumerate and simulate a limited number of TMs $M_i$ on the input of $M_{opt}$ and accept if any $M_i$ accepts, after verifying that this $M_i$ accepts correctly. Crucial to the strategy are: (1) the existence of a nondeterministic machine $N_c$ which is used to verify that $M_i$ accepts correctly, and (2) the existence if (*) fails of an $M^*$ which can efficiently simulate $N_c$.

Assume that (*) does not hold. That is, there exists some deterministic machine $M^*$ for \texttt{coBHP} such that for any non-halting $\langle N',x'\rangle$, there exists some polynomial $p_{N',x'}$ such that $M^*$ accepts $\langle N',x',1^t\rangle$ in at most $p_{N',x'}(t)$ steps for all $t$. Assume some enumeration $M_1, M_2,\ldots$ of deterministic machines and assume that the first machine $M_1$ in the enumeration is a standard $2^{cn}$-time machine accepting \texttt{coBHP} by brute force. Consider the following deterministic machine $M_{opt}$:

\begin{enumerate}
\item
Input $y=\langle N,x,1^t\rangle$ for the \texttt{coBHP} problem (let $n = |y|$);

\item
For each $\tau = n, n+1, \ldots$, run all machines $M_1, \ldots, M_n$ on $y$ within $\tau$ steps:

If $M_1$ terminates and accepts $y$, then accept $y$ and halt;

If some $M_i$ accepts $y$, then accept $y$ and halt after verifying that:

\begin{quote}
(**) There is no instance of \texttt{BHP} of length $\le \tau$ such that $M_i$ wrongly accepts it in $\tau$ steps.
\end{quote}

\end{enumerate}
Clearly $M_{opt}$ accepts \texttt{coBHP} and otherwise does not halt.

The key idea is to reduce the problem of checking (**) to some halting problem in \texttt{coBHP}. Consider the execution of the following nondeterministic machine $N_c$:

\begin{enumerate}
\item
Input $x' = M_i$;

\item
For each $\tau' = 1, 2, \ldots$,

Guess $z$ ($=\langle N,x,1^t\rangle$) of length $\le \tau'$ and $w$ in $\{0,1\}^{\tau'}$;

If $M_i$ accepts $z$ within $\tau'$ steps and $w$ witnesses $z$ in \texttt{BHP}, then accept $x'$ and halt.

\end{enumerate}
Then $\langle N_c,M_i\rangle$ is non-halting if $M_i$ accepts \texttt{coBHP}. Furthermore, there is a polynomial $p_c(\tau)$ ($\approx\tau(\tau-1)/2$) independent from $x' = M_i$ such that:

\begin{enumerate}
\item
If $N_c$ on $M_i$ does not halt in $p_c(\tau)$ steps (i.e., $N_c$ on $M_i$ has no accepting path of length $\leq p_c(\tau)$), then (**), and
\item

$N_c$ on $M_i$ does not halt in $p_c(\tau)$ steps
iff $M^*$ accepts $\langle N_c,M_i,1^{p_c(\tau)}\rangle$ in $p_{N_c,M_i}(p_c(\tau))$ steps.
\end{enumerate}

Thus, by running $M^*$ on $\langle N_c,M_i,1^{p_c(\tau)}\rangle$, we can guarantee (**) (if it is indeed possible) in $p(\tau)$ steps for some polynomial determined by $M_i$.

Then the running time of $M_{opt}$ satisfies condition (1) of Def. \ref{defn}, which contradicts that \texttt{coBHP} has superpolynomial speedup.
\end{proof}
Interestingly, each problem identified by Chen and Flum \cite{ChenFlum2} as having the same complexity as \texttt{coBHP} under fixed parameter tractable reductions, such as the set of arithmetic statements $\phi$ with no proof of fewer than $t$ steps,  is also the resource-bounded version of a non-c.e. language.
\section{Conclusion}
Their result and the parallel between Fact \ref{factnonhalting} and condition (*) suggest a correspondence between known facts in computability theory and hypotheses in complexity theory. As another example, G\"{o}del demonstrated speedup in the length of proofs of arithmetic statements \cite{Godel}, and a corresponding conjecture in proof complexity is that there is speedup in the length of proofs of tautologies (no $p$-optimal propositional proof system).

This correspondence serves several purposes. First, it suggests statements such as (*) which are interesting in themselves. Second, the correspondence may be interpreted as (weak) evidence that there are superpolynomial speedups for accepting \textbf{coNP}-complete languages and for proof length for propositional proof systems as has been conjectured \cite{Krajicek}. Finally, we suspect that the validity of the corresponding statements such as Fact \ref{factnonhalting} and (*) are closely linked.\footnote{Fact \ref{factnonhalting} implies a very weak, model dependent speedup for $\texttt{coBHP}$ (for details see \cite{Monroe2009}). For $M$ accepting $\texttt{coBHP}$, let $S_M$ be the set of non-halting $\langle N,x\rangle$ such that $M$ accepts $\langle N,x,1^\infty\rangle$ in finite time, where the encoding is such that $M$ does not necessarily read the full input. By Fact \ref{factnonhalting}, $S_M$ does not include some non-halting $\langle N',x'\rangle$. Then for any $M$, there exists $M'$ such that $S_{M'}=S_M\cup \{\langle N',x'\rangle\}$ is a strictly larger set than $S_M$. This $M'$ avoids reading the full input in more cases than does $M$. This line of argument also holds for $\texttt{coBHP}$ with $N$ deterministic, and does not hinge on the fact that $\texttt{coBHP}$ is \textbf{NP}-complete.} Resource-bounded versions of noncomputable problems may misbehave by failing to have an optimal algorithm or proof system, just as their noncomputable counterparts misbehave by failing to have any algorithm or proof system at all.

To pursue this linkage, we can define a version of (*) which like Fact \ref{factnonhalting} is constructive. For $M$ accepting $\texttt{coBHP}$, let $E$ be the set of non-halting $\langle N,x\rangle$ for which $f(t)$ is polynomially bounded. Suppose there is an $M_E$ which accepts $E$. Then a constructive version of (*) is:\footnote{Joseph and Young \cite{JosephY85} and Wang \cite{Wang90} define $p$-productive languages where the productive function yields a single problematic input, whereas in Conjecture \ref{constructive} an infinite family of inputs is produced.}
\begin{conjecture}\label{constructive}
The $\langle N',x'\rangle$ for $M_E$ predicted by Fact \ref{factnonhalting} satisfies (*) for $M$.
\end{conjecture}

The condition (*) was motivated by our suspicion that the existence of a polynomial time $M$ accepting $\texttt{coBHP}$ would violate the information constraint imposed by the noncomputability of the halting problem. More precisely:
\begin{conjecture}
If there exists $M\in \textbf{P}$ accepting $\verb"coBHP"$, then $M$ can be modified to accept all non-halting $\langle N,x\rangle$.
\end{conjecture}
More broadly, we wonder whether the obstacle to the existence of a TM $M$ which acts contrary to various widely believed complexity hypotheses is that $M$ could be modified to perform a related task known to be noncomputable. For instance, it is curious that arithmetic is undecidable only if it incorporates multiplication, and that this fact has not been used to say anything about the complexity of integer multiplication (which may have a slight speedup) or the inverse operation of factorization.

\bibliographystyle{model1b-num-names}
\bibliography{equivalence}
\end{document}